\documentclass[fleqn,usenatbib]{mnras}

\usepackage{newtxtext,newtxmath}

\usepackage[T1]{fontenc}
\usepackage{ae,aecompl}

\usepackage{graphicx}	
\usepackage{amsmath}	
\usepackage{amssymb}	
\usepackage{multicol}
\usepackage{graphicx}

\pdfoutput=1

\title[Galaxy Morphologies with Deep learning]{Improving galaxy morphologies for SDSS with Deep Learning}

\author[H. Dom\'inguez S\'anchez et al.]{
H. Dom\'inguez S\'anchez,$^{1, 2}$\thanks{E-mail: helenado@sas.upenn.edu}
M. Huertas-Company,$^{1, 2, 3}$
M. Bernardi$^{1}$, D. Tuccillo$^{2, 4}$ \and
and J. L. Fischer$^1$\\
$^{1}$Department of Physics and Astronomy, University of Pennsylvania, 209 South 33rd  Street, Philadelphia, PA 19104, USA\\
$^{2}$LERMA, Observatoire de Paris, PSL Research University, CNRS, Sorbonne Universit\'es, UPMC Univ. Paris 06,
F-75014 Paris, France\\
$^{3}$University of Paris Denis Diderot, University of Paris Sorbonne Cit\'e (PSC), 75205 Paris Cedex 13, France\\
$^{4}$MINES Paristech, PSL Research University, Centre for Mathematical Morphology, Fontainebleau, France\\
}

\date{Accepted XXX. Received YYY; in original form ZZZ}

\pubyear{2015}

\begin{document}
\label{firstpage}
\pagerange{\pageref{firstpage}--\pageref{lastpage}}
\maketitle

\begin{abstract}

 We present a morphological catalogue for $\sim$ 670,000 galaxies in the Sloan Digital Sky Survey  in two flavours:  T-Type, related to the Hubble sequence, and Galaxy Zoo 2 (GZ2 hereafter) classification scheme. By combining accurate  existing visual classification catalogues with machine learning, we provide the largest and most accurate morphological catalogue up to date. The classifications are obtained with Deep Learning algorithms using Convolutional Neural Networks (CNNs). 

We use  two visual classification catalogues,  GZ2  and \citet{Nair2010}, for training CNNs with colour images in order to obtain T-Types and a series of GZ2 type questions (disk/features, edge-on galaxies, bar signature,  bulge prominence, roundness and mergers). We also provide an additional probability enabling a separation between pure elliptical (E) from S0, where the T-Type model  is not so efficient. For the T-Type, our results show smaller offset and scatter than previous models trained with support vector machines. For the GZ2 type questions, our models have large accuracy  (\mbox{> 97\%}), precision and recall  values  (\mbox{> 90\%}) when applied to a test sample with the same characteristics as the one used for training. The catalogue is publicly released with the paper.

\end{abstract}




\section{Introduction}
\label{sect:intro}

Since the beginning of the last century, it is well known that galaxies exhibit a wide variety of morphologies. The first classification was done by  \cite{Hubble1926, Hubble1936}, dividing the galaxies into two broad types: galaxies with a dominant bulge component (also known as  early-type galaxies, ETGs) and galaxies with a significant disk component (late-type or spiral galaxies). The spiral galaxies are further divided into barred (with the presence of a bar shaped central structure) or unbarred, and ordered according to their spiral arms strength. The intermediate type between elliptical and spiral galaxies are called S0, while there is also a population of galaxies with irregular or distorted shapes.  According to this visual classification, a number can be assigned to each type of galaxy, which is known as the T-Type \citep{deVaucouleurs1963}.

Interestingly, morphology is very closely related to the stellar properties of the galaxies: in the local universe most elliptical galaxies show redder colours, larger masses, higher velocity dispersions and older stellar populations than spiral galaxies,  which are mostly gas rich star-forming systems with high rotation velocities (e.g., \citealt{Roberts1994,Blanton2009,Pozzetti2010} and references therein). It is also well known that both the structural and the intrinsic properties of galaxies undergo a significant evolution across cosmic time (e.g., \citealt{Wuyts2011},  \citealt{Huertas-Company2013}, \citealt{Huertas2015}, \citealt{Barro2017}). Understanding how morphology  relates to  all these other properties and in which way they affect galaxy assembly is one of the major challenges of present day astronomy.

It is, therefore, crucial to have accurate galaxy morphological classifications for large samples. Morphological classification has traditionally been  done by eye. However, this presents two major problems: first, it is not obvious how to categorise galaxies into one of each subclass, since there is a smooth transition between each  T-Type. This effect is even more evident at high redshift where, in addition to the poorer image quality, important structural changes and transitions between morphological types are taking place (e.g. \citealt{Huertas2015}). Second, visual classification is an incredible time-consuming task.  This is an enormous disadvantage in the era of big data, when extremely large surveys (such as SDSS, \citealt{Eisenstein2011}, Dark energy Survey, \citealt{DES2016} or EUCLID, \citealt{Racca2016}) release images for millions of galaxies. Visual classification does become a real impossible task.

One smart way to overcome the problem of visual classification for large amounts of data  was the Galaxy Zoo project\footnote{https://data.galaxyzoo.org/}, where ``science citizens" volunteered to classify galaxies through a user-friendly web interface. The first approach was a very simple classification into three types (ETGs, spirals or mergers) but, given the success of the project, a more complex classification system, GZ2, was proposed  in\citealt{Willett2013}. However, galaxy classifications made by amateur astronomers,  which is a difficult task even for professionals, has its caveats. For example, features such as bars are only selected when the bar is obvious and the volunteers tend to choose intermediate options when available (e.g. prominence of bulge, roundness, etc.).  There is also a large number of galaxies with uncertain classifications caused by the disagreement between classifiers.

Automated classifications using a set of parameters that correlate with morphologies (e.g. concentrations, clumpiness, asymmetries, Gini coefficients, etc.) have also been attempted \citep{Abraham1996, Conselice2000, Lotz2008}. A generalisation of that approach, using an n-dimensional classification with optimal non-linear boundaries in the  parameter space, was proposed in \cite{Huertas2011}.

A natural step  forward is to take advantage of the recently popular Deep Learning algorithms, which do not require a pre-selected set of parameters to be fit into the model but are able to automatically extract high-level features at the pixel level. In particular, CNNs have been proven very successful  in the last years for many different image recognition purposes: manuscript numbers, facial identification, etc. (e.g., \citealt{Ciresan2012, Russakovsky2014}).
CNNs have also been used for morphological classification of galaxies, with a high success rate.  The use of these automated classification algorithms has been possible thanks to a series of advances in the last few years: the  existence of large number of classified objects needed for the training (thanks to Galaxy Zoo project, in particular), the available computing power and a new set of techniques (e.g. rectified linear units -ReLUs-  \citealt{icml2010_NairH10} or dropout regularization, \citealt{Hinton2012}, \citealt{Srivastava2014}), as well as open source codes which facilitate the task. For example \cite{Huertas2015} applied  CNNs  to  classify ~50,000 CANDELS \citep{Grogin2011ApJS..197...35G,Koekemoer2011} galaxies into  five groups (spheroid, disk, irregular, point source and unclassifiable.). They obtained zero bias, $\sim$ 10\% scatter and less than 1\% of misclassification.  The CNN model presented in \citet[D15 hereafter]{Dieleman2015}, was able to reproduce the GZ2 classification with large accuracy for galaxies with certain classifications. However, one problem with this work is that all biases from GZ2 visual classifications are included; i.e., all the GZ2 catalogue is used for training the models, even galaxies with uncertain classifications.

We follow up that work and create an improved version of the GZ2 catalogue by  training our models only  with galaxies with very robust GZ2 classification. We also simplify the galaxy decision tree by giving only one probability value for each question (see section \ref{sect:GZ2}). In addition, we complement the GZ2 classification scheme with a T-Type, trained with the visually classified catalogue from \citet[N10 hereafter]{Nair2010}.  The T-Type is an extremely useful parameter for morphological classification because it gives  information about the relative importance of the bulge and disk components by one single number.  We also use the N10 catalogue to provide a model to separate pure E from S0's and an alternative bar classification to the GZ2 based one. We provide all these values for the sample of $\sim$ 670,722 galaxies in the  Sloan Digital Sky Survey (SDSS) Data Release 7 (DR7)  Main Galaxy Sample \citep{Abazajian2009} with r-band Petrosian magnitude limits $14 \le m_r \le 17.77$ mag published by \citet[see Section \ref{sect:sample}]{Meert2015, Meert2016}. This is a significant increase in the number of classified galaxies compared to  similar available morphological catalogues (almost three times larger than the  GZ2  and $\sim$ 50  times larger than the N10).

The paper is organised as follows: in Section \ref{sect:data} we introduce the data sets used for training and testing our models, as well as the sample for which the catalogue described in this paper is released. In Section \ref{sect:models} we describe the Deep Learning model and its network architecture. In Section \ref{sect:GZ2}  and \ref{sect:Nair-models} we present the methodology and results of our models trained with the GZ2 and the N10 catalogues, respectively. Finally Section \ref{sect:catalogue} details the content of our morphological catalogue and  Section \ref{sect:conclusions}  summarises our main results.

\section{Data sets}
\label{sect:data}

To carry out this work we have benefited from a series of morphological galaxy catalogues, which we use to train and test our Deep Learning models. In this section, we describe the datasets used for training and testing, as well as the final sample to which we apply our models and for which we release our catalogue.

\subsection{Catalogues used for training the models }
\label{sect:data-train}
\subsubsection{The Galaxy Zoo 2 catalogue}
The GZ2 is a public catalogue for $\sim$ 240,000 galaxies  (m$_{r}$ < 17 mag,  z < 0.25) of  the  SDSS DR7 Legacy Survey, with classifications from volunteer citizens. The volunteers  have to answer a set of questions for each galaxy image. Depending on the answer, the user is directed to a different question  following the GZ2 decision tree. The GZ2 decision tree has 11 classification tasks with 37 possible responses (the number of  possible answers per question range from two to seven). We encourage the reader to refer  to  \citet[ W13 hereafter]{Willett2013} for a detailed description and, in particular, to Figure 1 for a better understanding of the GZ2 classification scheme, which will be  of significant importance throughout this work. The GZ2 catalogue includes number counts of votes and fractions for each answer (weighted and debiased, to correct from observational effects). We take advantage of the GZ2  catalogue for training our models on galaxy classifications similar to the GZ2 decision tree scheme. We base our analysis on weighted fraction values.  The  weighted fractions are calculated by correcting the vote fractions  with a  function  which down-weights classifiers in the tail of low consistency (see W13 for a detailed explanation). Classification bias corrections  have been derived in W13 (and refined in a recent work by \citealt{Hart2016}). The debiased fractions account for changes in the observed morphology as a function of redshift,  independent of any true evolution in galaxy properties. The debiased values contain additional information which is not actually included in the images. Therefore, we prefer to restrict our analysis to weighted fractions.  Weighted factions are used exclusively hereafter and we will refer to them as P$_{task}$, where \textit{task} is the particular question being discussed.

\subsubsection{Nair et al. 2010 catalogue}
The \cite{Nair2010} is a  catalogue based on visual classifications of monochrome g-band images by an expert  astronomer for 14,034 galaxies in the SDSS-DR4 in the redshift range 0.01 < z < 0.1 down to an apparent extinction-corrected limit of m$_g$ < 16 mag. The data include RC3 T-Types, as well as the existence of bars, rings, lenses, tails, warps, dust lanes, etc. The N10 catalogue provides a detailed bar classification, which distinguishes between strong, intermediate and weak bars (plus additional features and combinations of them).  We use  the N10 catalogue  to train our models for T-Type classification, for a complementary  bar classification, and  to separate pure E form S0 galaxies.

\subsection{Catalogues used for testing the models }
\label{sect:data-test}
To study the performance of our models,  we combine tests on the catalogues used for training (described in Section \ref{sect:data-train}) with tests on available catalogues which are not used in the training process.

\subsubsection{Huertas-Company et al. 2011 catalogue}
In order to test how our T-Type classification compares with  previous automated classifications, we use \cite{Huertas2011} catalogue. This dataset contains an  automated morphological classification in 4 types (E, S0, Sab, Scd) based on support vector machines of $\sim$~670,000 galaxies from the \citet{Meert2015} SDSS DR7 sample. Each galaxy is assigned a probability  of being in the four morphological classes instead of assigning a single class. We then transform these probabilities into T-Types by using equation 7 from \citet{Meert2015}.

\subsubsection{Cheng et al. 2011}
 The  \cite{Cheng2011}  catalogue  consists of 984 non-star-forming SDSS galaxies with apparent sizes >14 arcsec and is focused on making finer distinctions between ETGs. It includes a visual classification plus an automated method to closely reproduce the visual results. Galaxies are divided into three bulge classes  by the shape of the light profile in the outer regions, roughly corresponding to Hubble types E, S0 and Sa. We use \citet{Cheng2011} catalogue to test the ability of our models to properly separate S0/Sa from pure E galaxies (see section \ref{sect:Ell/S0}).

\subsection{Parent sample of the morphological catalogue presented in this work}
\label{sect:sample}
The catalogue released along with this paper is based on the sample  described in \cite{Meert2015, Meert2016} in order to take advantage of the quality of processed data available for these galaxies. The Meert et al. catalogue contains 2D decompositions in the $g$, $r$, and $i$ bands for each of the de~Vaucouleur's, S{\'e}rsic, de~Vaucouleur's  + exponential disk  and S{\'e}rsic + exponential disk models.  As discussed in a series of papers (\citealt{Bernardi2013, Meert2015, Fischer2017, Bernardi2017b} and references therein) the SDSS pipeline photometry underestimates the brightnesses of the most luminous galaxies. This is mainly because (i) the SDSS overestimates the sky background and (ii) single or two-component S{\'e}rsic-based models fit the surface brightness profile of galaxies better than the de~Vaucouleur's model used by the SDSS pipeline, especially at high luminosities.  In addition to having substantially improved photometry, stellar masses for the objects in this catalogue have recently been added \citep{Bernardi2017a}.  Therefore, further augmenting this rich data set with morphological information represents a significant added-value. The reader can refer to \citet[M15 hereafter]{Meert2015} for a more detailed description of the sample selection.  Once trained and tested, we apply our morphological classification models to  all galaxies in that dataset. For each galaxy,  we provide a  probability for each of the questions listed in Table \ref{tab:questions},  based on GZ2 catalogue.  We use the N10 catalogue to derive a T-Type and also a probability value of being S0 versus E (to better separate galaxies with T-Type $\leq$ 0), plus an  additional bar classification. In section \ref{sect:catalogue} we summarise the catalogue content and give  advise on how to properly use it.

\section{Deep Learning morphological classification model}
\label{sect:models}
In this work, we apply Deep Learning algorithms using CNNs to morphologically classify galaxy images. Deep Learning is a methodology which automatically learns and extracts the most relevant features (or parameters) from raw data for a given classification problem through a set of non-linear transformations. The main advantage of this methodology  is that no pre-processing needs to be done: the input to the machine are the raw RGB cutouts for each galaxy. The main disadvantage is that, given the complexity of extracting and optimizing the features and weights in each layer, a large number of already classified images need to be provided to the machine. Fortunately, as explained in Section \ref{sect:data}, there is a wealth of morphological catalogues in the literature overlapping our dataset, which we can use for training and testing our model performance.

\subsection{Network architecture}

\begin{figure*}
\centering
\includegraphics[width=0.9\textwidth]{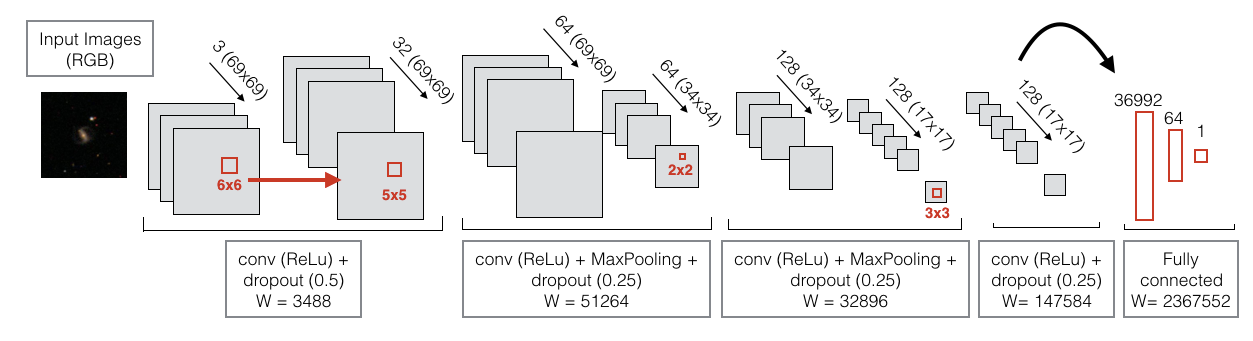}
\caption{Network architecture used for training the models, consisting on four convolutional layers and a fully connected layer, as explained in the text. The number of weights at each level (W)  are  indicated.}
\label{fig:network} 
\end{figure*}

Given the high rate of success of previous works using CNNs for visual classification of galaxies \citep{Huertas2015, Dieleman2015}, we adopt a similar (but not identical) CNN configuration. Testing the performance of different network architectures is beyond the scope of this paper, and we use the same input images and CNN configuration for each classification task. We use the  \texttt{KERAS} library\footnote{https://keras.io/}, a high-level neural networks application programming interface, written in  \texttt{Python}.

The input to our machine are the RGB cutouts downloaded from the SDSS DR7 server\footnote{http://casjobs.sdss.org/ImgCutoutDR7} in $jpeg$ format, with  424$\times$424 pixels of  0.02$\times$R$_{90}$  arcsec in size (per pixel, where R$_{90}$ is the Petrosian radius for each galaxy). The algorithm reads the images which are down-sampled into (69, 69, 3) matrices, with each number representing the flux in a given pixel at a given filter. Down-sampling the input matrix is necessary to reduce the computing time and to avoid over-fitting in the models. The flux values are  normalised to the maximum value in each filter for each galaxy. The network architecture, represented in Figure \ref{fig:network}, is composed of four convolutional layers with squared filters of different sizes (6, 5, 2 and 3, respectively) and a fully connected layer. Dropout is performed after each convolutional layer to avoid over-fitting, and a 2$\times$2 max-pooling follows the second and third convolutional layers. The number of weights  in each layer -before dropout- are also indicated. The output of the fully connected layer is a single value, which has different meanings for each model (see Sections \ref{sect:GZ2} and \ref{sect:Nair-models}). 

We train the models in  binary classification mode for GZ2 based questions and in regression mode for the T-Type values. The output of the models trained in binary classification ranges from 0 to 1, and it can be interpreted as  the probability of being a positive example (example labelled as Y=1 in our input matrix). The output of the T--Type model trained in regression mode ranges from -3 to 10, and the returned value is directly the T-Type. We use 50 training epochs, with a batch size of 30 and  (usually) a \textit{learning rate}  of 0.001. We tested the effect of using different \textit{learning rate} values for questions which were more difficult to train (e.g., bars, bulge prominence and roundness). In the training process, we perform  \textit{data augmentation}, allowing the images to be zoomed in and out (0.75 to 1.3 times the original size), rotated (within 45 degrees), flipped and shifted  both vertically and horizontally (by 5\%). This ensures our model does not suffer from over-fitting since the input is not the same in every training epoch.

\section{Galaxy Zoo 2 based models}
\label{sect:GZ2}
In this Section, we explain in detail the training methodology and the results obtained for the GZ2  based models listed in Table \ref{tab:questions}.
\subsection{Training methodology}
\label{sect:train}

In this work we use the W13 catalogue  for training our GZ2-based models. In  D15, a CNN  able to reliably predict various aspects of GZ2 galaxy morphology directly from raw pixel data was presented.  While their objective was to reproduce the whole GZ2 catalogue, we aim to provide an improved version of the GZ2 classification. In D15 the goal was to predict probabilities for each answer simultaneously solving a regression problem, while we train each question independently using a binary mode classification algorithm.  Our main difference with respect to D15  approach is that we  only use for the training of each question galaxies with low uncertainties in the GZ2 classification. This allows the model to better identify the important features for each task and to obtain a  more evident classification for galaxies for which the GZ2 classification was  uncertain (see Section \ref{sect:catalogue}). 

We do not try to reproduce the whole GZ2 decision tree, but we restrict our analysis to the questions belonging to the third tier. Questions in the lower levels of the classification tree are usually  classified by a smaller number of volunteers, reducing the statistics of robust samples, which is fundamental for training our models. Even though in the third tier, we do not address the spiral arm signature nor the bulge shape  questions (Q4 and Q9 in W13, respectively), since we believe  these tasks are too detailed for the resolution of our binned input images. The tasks included in this work are listed in Table \ref{tab:questions}.

\begin{table*}
\centering
\begin{tabular}{l|l|c| c | c| c |}
 Question  & Meaning &  N$_{votes}$ & N$_{certain}$  & N$_{pos}$  \\\hline
  Q1 & Disk/Features &  239728 (99\%) & 134475 (56\%)  & 28513  (21\%) \\
  Q2 & Edge-on disk &  151560 (63\%) & 123201 (81\%)  &  17631 (14\%) \\
  Q3 & Bar sign & 117262 (48\%)  & 76746 (65\%)  & 6595 (8\%) \\
  Q4 & Bulge prominence & 117245 (49\%) & 49345 (42\%)  & 27185 (55\%)\\
  Q5 & Cigar shape & 180223 (75\%) &124610 (70\%) &  28230 (23\%)\\
  Q6 & Merger  signature &  239669 (99\%) & 110079 (46\%) & 1399 (1\%) \\
\end{tabular}
\caption{\label{tab:questions} Questions from the GZ2 scheme addressed in this work (note that question numbers do not correspond to the ones in Table 2 from W13). Also shown the total number (and fraction) of galaxies with enough votes  in GZ2 to be used in the training ($>$ 5,  N$_{votes}$), the number of \textit{certain} galaxies (N$_{certain}$) which fulfil our requirement for being used in the training ($a(p) > 0.3$, see  text for a detailed explanation for each question) and the number of positive examples for each question (N$_{pos}$, e.g.  number of galaxies with a bar signature in Q3).  The percentages are  derived from the parent sample of the previous column (i.e., the fraction of N$_{certain}$ is the number of \textit{certain} galaxies divided by the number of galaxies with enough votes).}
\end{table*}

The fact that GZ2 classifications are based on the answers of citizens, who may not have any background on galaxy images, has some inconveniences.    One of the most troublesome tasks is the identification of bar signatures: only the most prominent bars have a high probability of being identified as such, while the weaker features are hardly recovered. For example,  only 50\% of the weak bars identified by N10 have P$_{bar}$ $>$ 0.5 in the GZ2 catalogue (P$_{bar}$ $>$ 0.5  is the threshold used in \citealt{Masters2011} to select GZ2 barred galaxies). Mergers are also difficult to identify simply by eye, and the sample of galaxies with large P$_{merger}$  in GZ2 is heavily contaminated by projected pairs (see \citealt{Darg2010, Casteels2013}).   On the other hand, the advantage of the GZ2 classification is  that there are sufficient statistics  to investigate and quantify these issues. 


\begin{figure}
\centering
\includegraphics[width=0.5\textwidth]{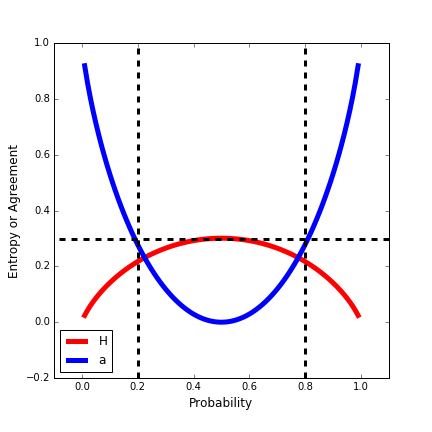}
\caption{Entropy ($H(p)$, red line) and agreement ($a$(p), blue line) versus probability for binary questions, where $P_1$+$P_2$=1. The dashed line marks the limit used throughout the paper to consider a galaxy in GZ2 as  \textit{robust} classification: $P_i$ < 0.2 or $P_i$ > 0.8, roughly corresponding to $a(p)$ $\geq$ 0.3.}
\label{fig:H-a} 
\end{figure}

 When the answer for a particular question is not  obvious for the volunteers, the vote fractions take intermediate values, meaning that the GZ2 classification  for those cases  are rather uncertain (see Table \ref{tab:questions}). Following  D15, we  quantify the agreement between classifiers,  $a(p)$:

\begin{equation}
a(p) =1- \frac{H(p)}{log(n)}
\label{Eq:a(p)}
\end{equation}

where $H(p)$ is the entropy of a  question with  $n$  possible answers and probability $p(x_i)$ for answer $i$:

      \begin{equation}
      H(p) =-\sum_{i}^{n} p(x_i)logp(x_i)
      \label{Eq:H}
      \end{equation}

The meaning of  $a(p)$ is a measurement of how consistent a classification is, for all the participants that answered that question. In Figure \ref{fig:H-a} we show the behaviour of the two functions, $H(p)$ and $a(p)$, for a binary classification. Around 44\% of the galaxies in the GZ2 catalogue have an agreement lower than 0.3 for Q1, corresponding approximately to a probability between 0.2 - 0.8 for a binary question (see Table \ref{tab:questions}). This complicates the usage of   the GZ2 catalogue in scientific studies. Another problem is the number of classifiers that have answered a particular question, i.e., the minimum number of votes needed to consider a classification as reliable.

Our methodology consists in only using  galaxies with a very robust classification in GZ2  for training each  question: we require $P$ > 0.8 in one of the two possible answers (these limits are relaxed to 0.7 for questions where the statistic is limited) and a minimum of five vote counts (at least five people have answered that question) in order to use a galaxy in our training sample. This removes noisy galaxies, which are difficult to classify by humans, and allows the model to more rapidly converge. The price to pay is that we have fewer galaxies to train in every question, as can be seen in Table \ref{tab:questions}. 

In addition,  instead of allowing more than two answers for some questions, as in the original GZ2 scheme (e.g.  the bulge prominence question has four possible outputs: \textit{no bulge, just noticeable, obvious, dominant}), we train our models in  binary classification mode, i.e., only positive or negative examples are provided.  The loss function used throughout this work for binary classification tasks is \texttt{binary-crossentropy} with \texttt{adam}  optimizer and  \texttt{sigmoid} activation. Since the output of our model is a probability distribution, that number can be interpreted as the degree of, e.g.,  bulge importance or roundness.  

To summarise, there are three main differences in our methodology compared to D15:

\begin{itemize}
\item We train each question individually, i.e., we use one model for obtaining each of the parameters contained in the catalogue.
\item We use ONLY robust classifications for training our models (more than five votes and $a(p)$ $\geq$ 0.3).
\item We train the models in binary mode, not in regression mode.

\end{itemize}

 \begin{figure*}
\centering
\includegraphics[width=0.9\textwidth]{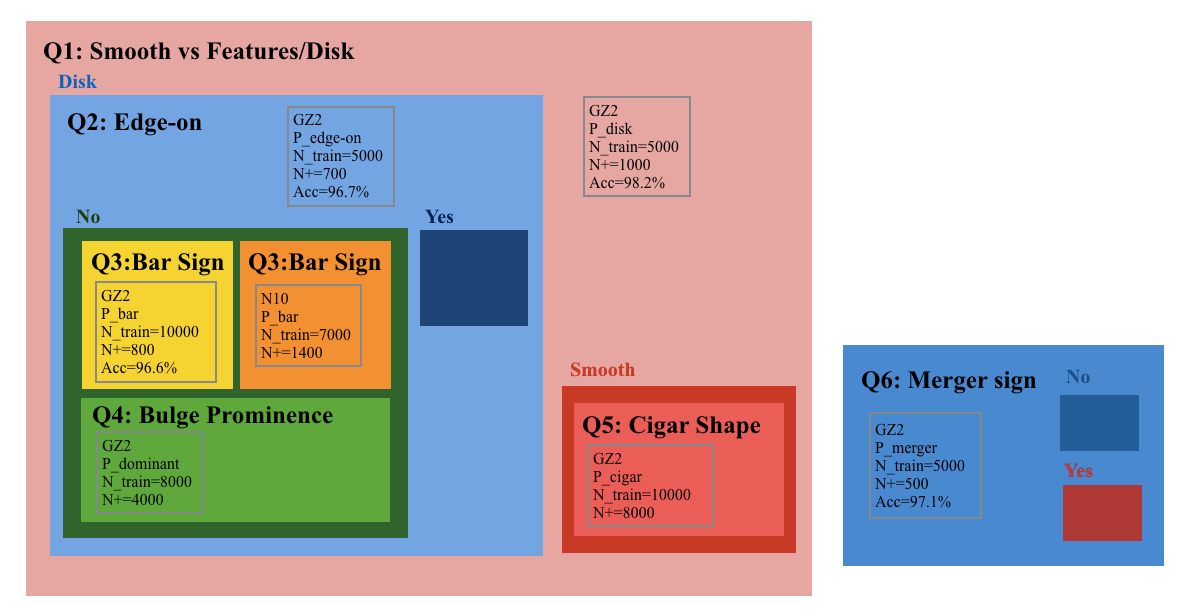}
\caption{Scheme for our classification of GZ2 type questions. Each box represents a model, with some characteristics framed in grey (from top to bottom: the catalogue used for training,  the output of the model, the number of galaxies used in the training, the number of positive examples in the training and the average accuracy - when its computation is feasible). Each box contains additional boxes representing  the two possible answers of the model, which may, at the same time, contain additional boxes representing questions trained for that particular subset of galaxies (e.g., the bar classification is only trained with non edge-on disk galaxies.) }
\label{fig:scheme} 
\end{figure*}

Figure \ref{fig:scheme} shows the classification scheme for the GZ2 type questions. Here we describe in detail some particularities on the training for each  question in Table \ref{tab:questions}:
 
\begin{itemize}
 \item {\bf Q1 - Disk/features:}
This question classifies smooth galaxies versus galaxies with the presence of disk or features. It is the first question in the GZ2 classification scheme and has, therefore, been answered by all the participants. Only 20 galaxies in the whole catalogue have less than five votes (adding the smooth and the disk/feature votes), meaning that statistics is not an issue when training this question. However, only  56\% of them have a \textit{certain} classification, i.e., satisfy the requirement of having  $P_{smooth}$   > 0.8 or $P_{disk}$  > 0.8, of which $\sim$ 21\% are classified as \textit{disk/features}. We use 5000 galaxies in the training ($N_{train}$). For this particular task, this number of galaxies is enough for the models to converge (i.e., setting $N_{train}$=10000 does not improve the model performance). We consider as positive examples  galaxies with  $P_{disk}$ > 0.8. The output of the model is the probability of galaxies having disk or features, $P_{disk}$.  \\
 \item {\bf  Q2 - Edge-on galaxies:}
This question belongs to the second level of the GZ2 classification scheme (only participants who choose the \textit{disk/features} path were asked this question) and $\sim$ 63\%  of the galaxies have $>$ 5 votes. However, this is a pretty evident question and $\sim$ 81\%  of the galaxies have a \textit{certain} GZ2 classification (P > 0.8 in one of the two answers), of which only 14\% are edge-on (positive examples). To overcome the small number of positive examples, we use balanced weights  (i.e., each instance of the smaller class - edge-on galaxies - contribute more to the final loss, whereas the larger class - non edge-on galaxies -  contribute less). The output of the model, trained with  $N_{train}$=5000 galaxies, is $P_{edge-on}$.\\
 \item {\bf  Q3 - Bars:}
This question belongs to the third level of the GZ2 classification scheme (only participants who choose the \textit{disk/features} and \textit{no edge-on} path were asked this question), reducing the sample of  galaxies which have at least five votes to $\sim$ 48\%. The fraction of them having P > 0.8 in one of the two answers is  $\sim$ 65\%, of which only 8\% are barred galaxies (positive examples). The small number of barred galaxies complicates the training, which we overcome by increasing the training sample ($N_{train}$=10000) and using balanced weights. The output of the model is the probability of having bar sign, $P_{bar}$.\\
 \item {\bf Q4 -  Bulge Prominence:}
 This question also belongs to the third level of the GZ2 classification scheme (only participants who choose the \textit{disk/features} and \textit{no edge-on} path were asked this question), reducing the sample of  galaxies which have 5 votes to $\sim$ 49\%. In the GZ2 classification, this questions has four possible answers (\textit{no bulge, just noticeable, obvious} or \textit{dominant}). The fraction of them having P > 0.7 in one of the  answers is  < 30\%, of which only 132 are bulge dominated. Requiring  $P_{dom}$+$P_{obvious}$ > 0.7, the fraction increases  to 42\%. Due to the scarce statistic and for simplicity reasons, we train the model related to this question in a binary classification mode: we consider as positive examples galaxies with obvious or dominant bulge ($P_{dom}$+$P_{obvious}$ > 0.7,   $\sim$ 55\% of the \textit{certain} sample) against galaxies with no bulge ($P_{no-bulge}$ > 0.7).  To obtain better results the \textit{learning rate} value used for training this question was set to 0.0001. The output of the model,  trained with $N_{train}$=8000, is $P_{bulge}$, i.e., the probability of having an obvious/dominant bulge. We  tested that this is the configuration which returns the best results. \\
   \item {\bf Q5 - Roundness:}
 This question  belongs to the second level of the GZ2 classification scheme (only participants who choose the \textit{smooth} option in Q1 were asked this question) and  $\sim$ 75\% of GZ2 galaxies have five or more votes. In the GZ2 classification, there are three possible answers  to this question (\textit{completely round, in between} and \textit{cigar shaped}) and  the fraction having P > 0.7 in one of the two answers is  $\sim$ 70\%, of which more than a half (63\%) are in the \textit{in between} category.  We proceed as in Q4 and  train the model related to this question in a binary classification mode: we consider as positive examples cigar shape galaxies (\mbox{$P_{cigar}$ > 0.7}) against completely round galaxies (\mbox{$P_{round}$ > 0.7}). To obtain better results the \textit{learning rate} value used for training this question was set to 0.0001.  The output of the model,  trained with $N_{train}$=10000, is   $P_{cigar}$, i.e. the probability of having a cigar shape instead of a round shape. \\
  \item {\bf  Q6 - Mergers:}
   This question belongs to the second level of the original GZ2 classification scheme. Although it is independent of the first answer on Q1, only users who answered \textit{yes} to the question \textit{Is there anything odd?}  are then directed to the next question (\textit{what is the odd feature?}), which has seven possible answers: merger, ring, arc/lens, distorted, irregular, dust lane or other. Only $\sim$7\% of the GZ2 galaxies have more than five counts in the merger answer, which limits the training sample. We choose a different approach to the  GZ2  scheme: we train a model in binary classification mode,  as we did with the previous questions. We consider as positive examples galaxies  with high probability of being merger  combined with a low probability of  no presenting anything odd ($P_{merger}$ > 0.7 and $P_{no-odd}$ < 0.45), against galaxies which are clearly non merger   ($P_{no-odd}$ > 0.9 and  $P_{merger}$ < 0.4  and at least 10 votes in the \textit{no-odd} answer). Since there are only $\sim$ 1400 clear merger examples, we use balanced weights. The output of the model,  trained with $N_{train}$=5000, is $P_{merger}$, i.e., the probability of presenting a merger signature. Given the scarce number of merger examples, this was the most challenging question to train in our models.  We leave for a forthcoming paper the use of simulated mergers for training a more curated model for merger identification.\\
 
\end{itemize}

\begin{figure*}
\setlength{\columnsep}{3pt}
\begin{multicols}{2}
    \includegraphics[width=\linewidth]{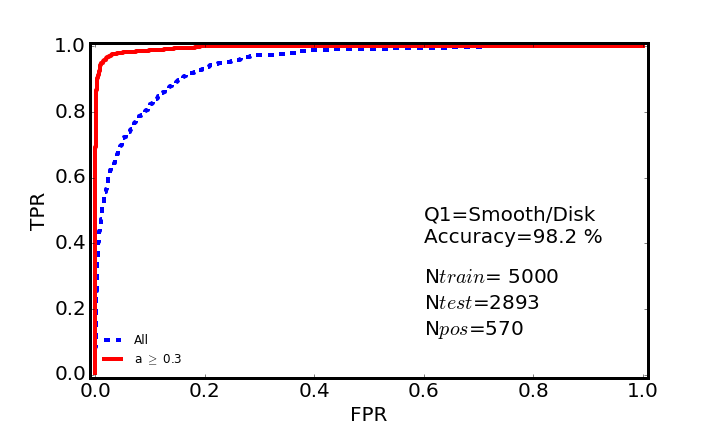}\par 
    \includegraphics[width=\linewidth]{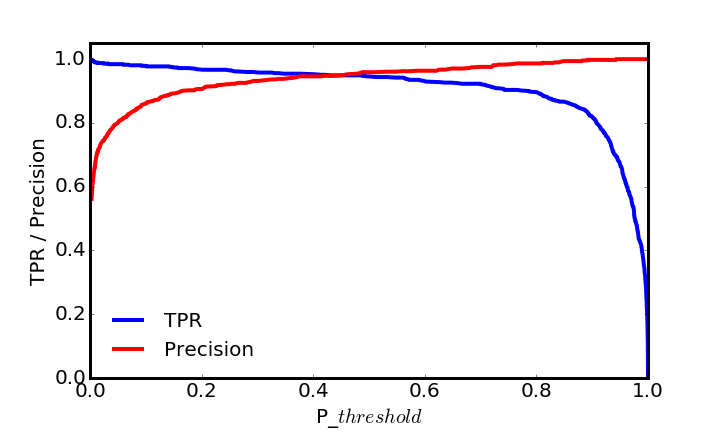}\par 
    \end{multicols}
 \vspace{-0.94cm}
\begin{multicols}{2}
    \includegraphics[width=\linewidth]{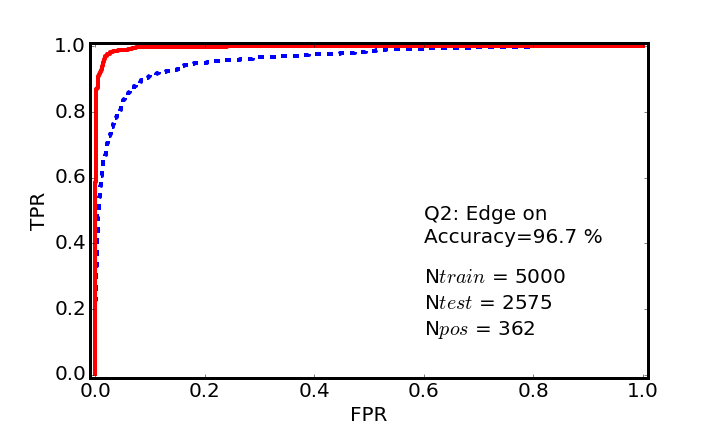}\par
    \includegraphics[width=\linewidth]{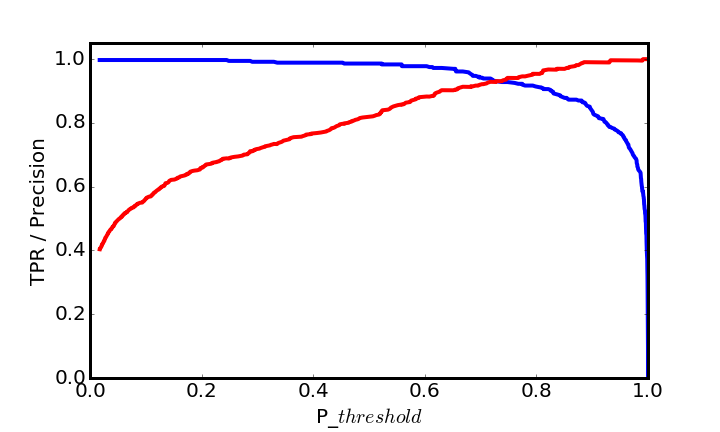}\par
\end{multicols}
 \vspace{-0.94cm}
\begin{multicols}{2}
    \includegraphics[width=\linewidth]{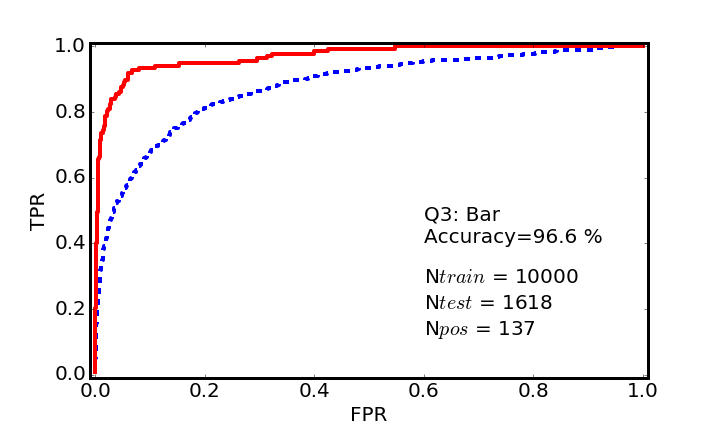}\par
    \includegraphics[width=\linewidth]{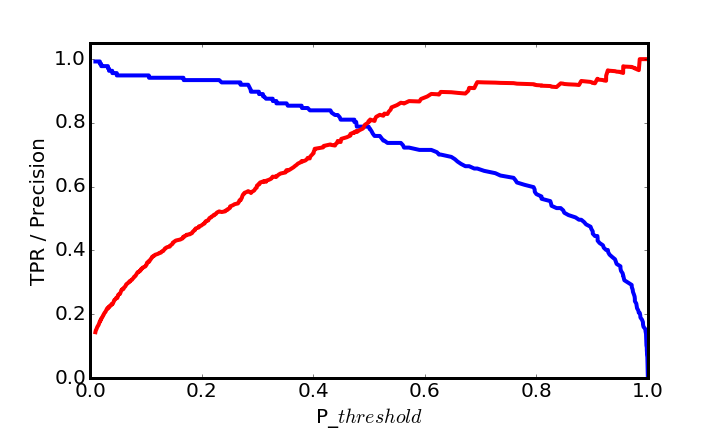}\par
\end{multicols}
 \vspace{-0.94cm}
\begin{multicols}{2}
    \includegraphics[width=\linewidth]{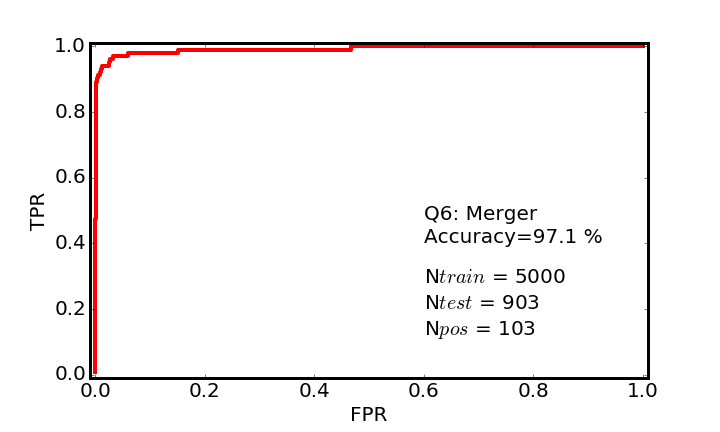}\par
    \includegraphics[width=\linewidth]{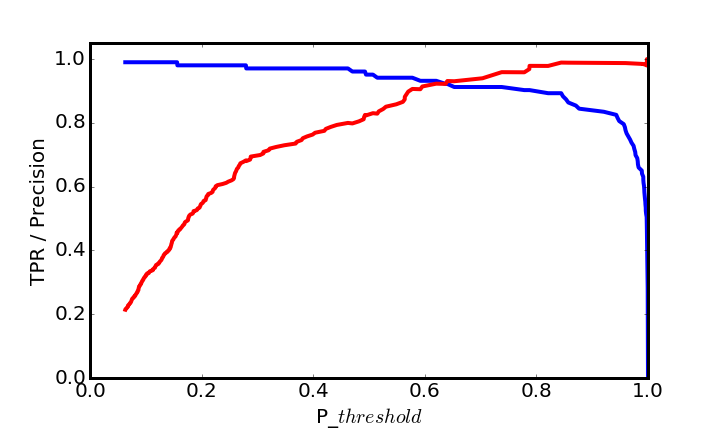}\par
\end{multicols}
\caption{ROC curves (left panels) and TPR, Precision values (blue and red lines, respectively) as a function of  $P_{thr}$ (right panels) for the four questions with only two possible answers in GZ2 (disk/features, edge on, bar and merger, from top to bottom). The red lines in the left panels show the results when applying the model to a test sample with the same characteristics as the one used for training ($a(p) \geq 0.3$ and at least five votes). The dashed blue line shows the ROC curve when applied to a test sample  without any cut in $a(p)$.  Also shown is the number of galaxies used in the training, the number of test galaxies, the number of positive test examples and the average accuracy. }
\label{fig:ROC}
\end{figure*}

\subsection{Testing the models}
In this Section, we detail the performance of our GZ2 based models when tested against a sample of robustly classified galaxies ($a(p)$ $\geq$ 0.3), comparable to the one used for training the models.  

\subsubsection{Questions with two possible answers}
\label{sect:GZ2-models}

In order to quantify the performance of our models for the questions with  only two possible answers in GZ2 (Q1, Q2, Q3, Q6), we use two standard methods from the literature: ROC curves and precision-recall versus probability threshold. 

A very common way to measure the accuracy of the models is the ROC curve of the classifier \citep{Powers2011}. This curve represents the false positive rate (FPR= FP/N,  i.e.,  the ratio between false positive and total negative cases)  versus true positive rate (TPR=TP/P, the ratio between true positive and total positive cases) for different  probability thresholds (P$_{thr}$). The better the classifier, the closer to the left $y$ axis and upper $x$ axis: i.e., it should maximise TP, and minimise FP values. A complementary way to test the model performance is the precision ($Prec$) and recall ($R$) scores (e.g.,  \citealt{Dieleman2015, Barchi2017}), which can be defined as follows:

\begin{align*}
Prec=\frac{TP}{TP+FP} ; \, \, \,  
R=\frac{TP}{TP+FN}=TPR  
\end{align*}

$R$, equivalent to the TPR, is a proxy of completeness, while $Prec$ is a purity (contamination) indicator. By choosing different P$_{thr}$ values  to consider a galaxy as a positive example, the $Prec$ and $R$ also vary. In Figure \ref{fig:ROC} we show these two tests  when applying our models to a control sample with similar characteristics  to the training sample (i.e., $a(p) \geq 0.3$ and  at least five votes) but not used for the training. The crossing point of the red and blue lines in the right panels is the  P$_{thr}$ value that optimises both the $Prec$ and $R$, but depending on the user purpose, one can vary the P$_{thr}$ to obtain a more complete or less contaminated sample. We tabulate precision  and recall values for P$_{thr}$=0.2, 0.5 and 0.8 for the 4 questions in Table \ref{tab:PRThr}.

The models have a  high success rate for all the questions, with total accuracy values defined as:
\[Acc=\frac{TP+TN}{(P+N)} \]

higher than 96\% and reaching 98\% for Q1 (see Table \ref{tab:PRThr}, Figure \ref{fig:ROC}). Also, for all the questions there is a P$_{thr}$ value for which both $Prec$ and $R$ > 0.9, except for Q3 (barred galaxies, which will be further discussed in \ref{sect:bar-nair}), for which the maximum is $\sim$ 0.8. This is related to the fact that bars are not easily identified by amateur astronomers, but also to the few positive barred examples in our training and testing samples (< 10\%), which causes the precision value to quickly decrease when few FP cases occur. If we consider the global accuracy of this question (fraction of the correctly classified galaxies), it reaches 96.6\%.

\begin{table}
\centering
\begin{tabular}{l|c| c| c|c| c|c| c|c|}
Question & Meaning             &  P$_{thr}$  & TPR    &  Prec.  &  Acc.     \\
                   &       			                 &   0.2   &  0.97  &  0.91      &              \\
         Q1    & Disk/Features   &    0.5  &  0.95   &  0.96      & 0.98   \\
      			  &			                		 &   0.8  &  0.90   &  0.99      &              \\
                                    
\hline

                   &       			                 &   0.2   &  1.00  &  0.67      &                  \\
         Q2    & Edge-on               &   0.5  &  0.99   &  0.83      & 0.97       \\
      			  &			                		 &   0.8  &  0.92   &  0.95      &                  \\
\hline

                   &       			                 &   0.2   &  0.93  &  0.48      &           \\
         Q3    & Bar sign               &   0.5  &  0.79   &  0.80      & 0.97 \\
      			  &			                		 &   0.8  &  0.58   &  0.92      &          \\
\hline

                   &       			                 &   0.2   &  0.98  &  0.54      &            \\
  Q6    & Merger signature       &   0.5  &  0.96   &  0.82     & 0.97   \\
      			  &			                		 &   0.8  &  0.90  &  0.97     &          \\

\end{tabular}
\caption{\label{tab:PRThr}Precision and recall (TPR) values for different $P_{thr}$  and average accuracy for the  questions which have two possible answers in GZ2 classification scheme. }
\end{table}

\begin{figure}
\centering
\includegraphics[width=0.5\textwidth]{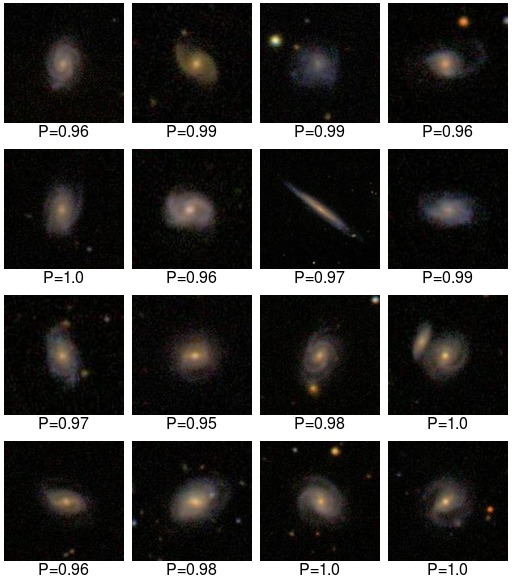}
\caption{Random examples of galaxies with a high probability of having disk/features according to our model,  shown  in each cutout. We note that the cutouts have been zoomed-in to the central third of the input images used by the CNN,  to better appreciate the detailed morphology. This  applies to all the cutouts shown throughout this work.}
\label{fig:disk}
\end{figure}

\begin{figure}
\centering
\includegraphics[width=0.5\textwidth]{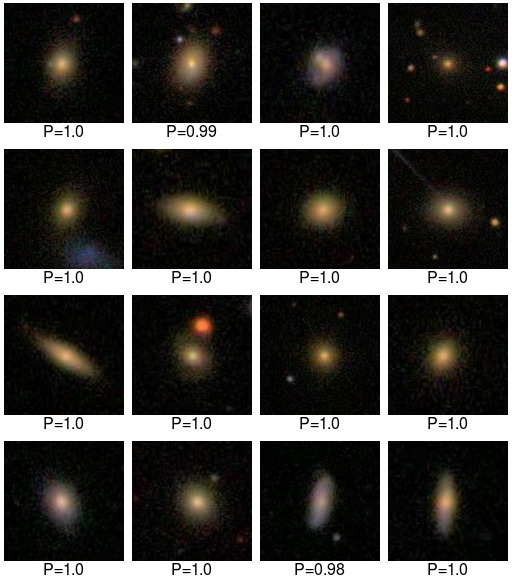}
\caption{Random examples of galaxies with a high probability of being smooth according to our model.}
\label{fig:smooth}
\end{figure}

\begin{figure}
\centering
\includegraphics[width=0.5\textwidth]{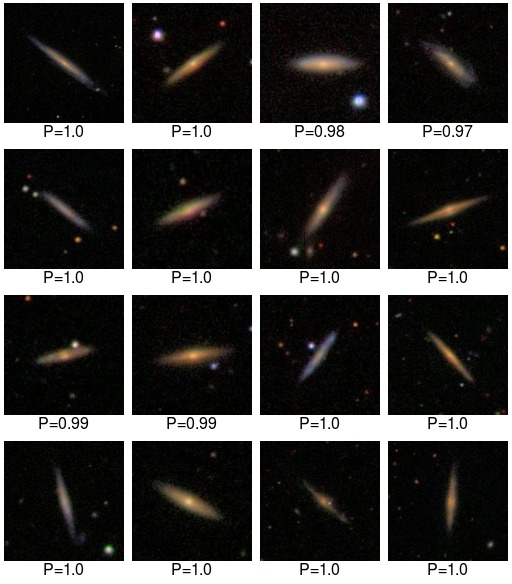}
\caption{Random examples of galaxies with a high probability of being edge-on according to our model.}
\label{fig:edge}
\end{figure}

\begin{figure}
\centering
\includegraphics[width=0.5\textwidth]{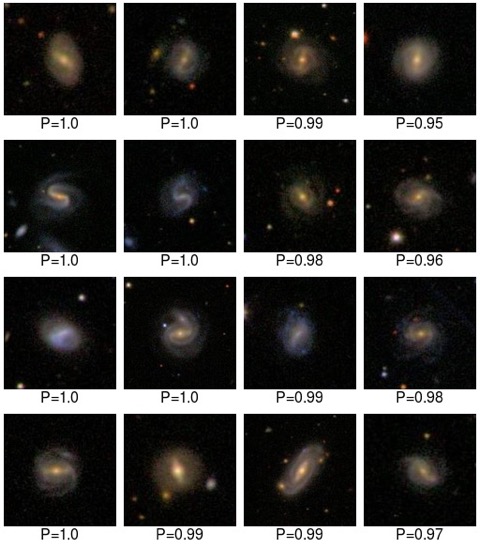}
\caption{Random examples of galaxies with a high probability of having bar signature according to our GZ2 model. Smooth and edge-on disks galaxies have been removed from the selection.}
\label{fig:bar}
\end{figure}

\begin{figure}
\centering
\includegraphics[width=0.5\textwidth]{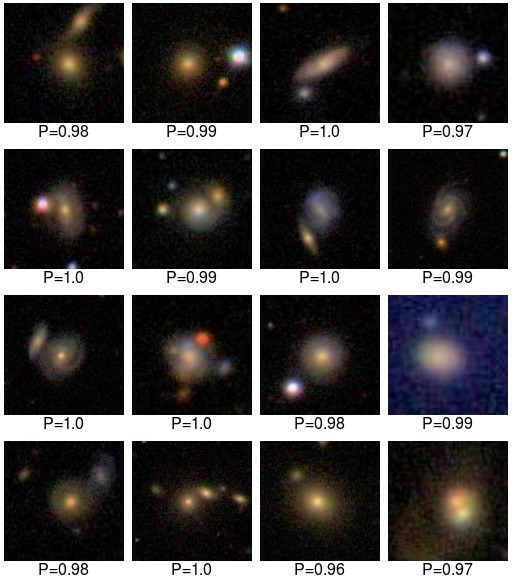}
\caption{\label{fig:merg-examples} Random examples of galaxies with high probability of showing merger signatures,  according to our model.}
\end{figure}

Finally, to visually inspect our models, we show some random examples of different galaxy types according to our classification:  disk/features, smooth, edge-on, barred and mergers (Figures \ref{fig:disk} - \ref{fig:merg-examples}).

\subsubsection{Questions with more than 2 answers}
\label{sect:2-answ}
\begin{figure*}
\setlength{\columnsep}{5pt}
\begin{multicols}{2}
    \includegraphics[width=\linewidth]{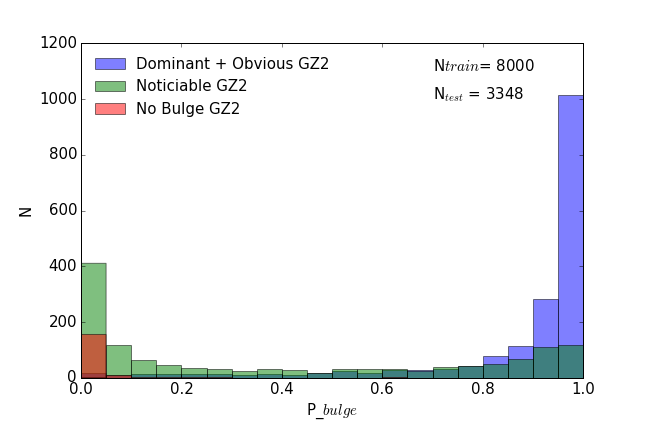}\par 
    \includegraphics[width=\linewidth]{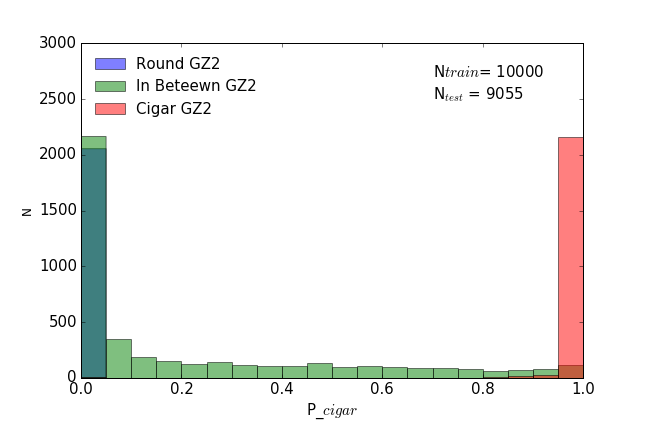}\par 
    \end{multicols}
\caption{ Probability distribution obtained by applying our models to a sample of well classified galaxies. The left panel shows the  probability of having a prominent bulge, while the  right panel shows the probability of being cigar shaped. Coloured bins represent galaxies with different GZ2 classifications, as stated in the legend. Also shown the number of galaxies used in the training and the number of test galaxies. Our classification is very efficient in separating the extreme cases in both questions.} \label{fig:P_distr}
\end{figure*}

The GZ2 scheme includes questions where more than two answers are possible, e.g., number of spiral arms (five possible answers) or prominence of the bulge (four possible answers). As already mentioned, we do not aim to reproduce the GZ2 classification scheme. In the case of questions with more than two possible answers, we have focused on the prominence of the bulge and the roundness of the galaxy. As explained in Section \ref{sect:train}, we train these questions on binary mode, discarding intermediate examples to avoid introducing noise in the training. 

Testing the behaviour of the models trained in this way by comparison with the GZ2 catalogue is not straightforward since we can not really define TP, TN, FP, FN values as we did for the binary mode questions. We can  test how well our derived probability distributions compare to the GZ2  classification for a sample  with similar characteristics to our training set (see section \ref{sect:train}). This is shown in Figure \ref{fig:P_distr}, for the probability of having a prominent bulge (P$_{bulge}$) and the probability of having cigar shape (P$_{cigar}$). 

The extreme cases for each question are clearly separated in the two models. For Q4, there is only a 2\% of FN (galaxies classified as bulge dominated in GZ2 which have P$_{bulge}$ < 0.4) and less than 0.1\% of  FP  (only three galaxies classified as having no bulge in GZ2 have P$_{bulge}$ > 0.5). For  galaxies classified as \textit{just noticeable bulge} in GZ2 the distribution is much wider, spanning all possible P$_{bulge}$ values, as expected for intermediate size bulges. There is a 6\% of those galaxies for which our model assigned a P$_{bulge}$ > 0.9 and 17\% with P$_{bulge}$ < 0.1. 

For question Q5, cigar shape versus round shape, the agreement between the GZ2 classifications and the model distributions is excellent, with less than 0.1\% of FP or FN (i.e., galaxies classified as round in the GZ2 with a high $P_{cigar}$ in our model and vice versa). The largest uncertainties are obtained for galaxies classified as \textit{in between} in GZ2, for which we find a 27\% with  P$_{cigar}$  <  0.1. This is  probably due to the fact that most GZ2 volunteers, when having an intermediate option, only choose the extreme cases (round or cigar) for the most evident examples.

\section{N10 based models}
\label{sect:Nair-models}
This work aims to provide the most complete and accurate morphological classification up to date  using Deep Learning models. For this reason, we complement the GZ2 classification with a T-Type model trained with the N10 catalogue, as well as an alternative bar classification. 

\subsection{T-Type model}
\label{sect:T-type model}

\begin{figure*}
\setlength{\columnsep}{1pt}
\begin{multicols}{3}
    \includegraphics[width=\linewidth]{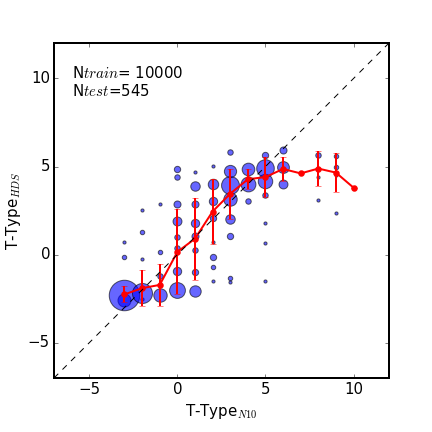}\par 
    \includegraphics[width=\linewidth]{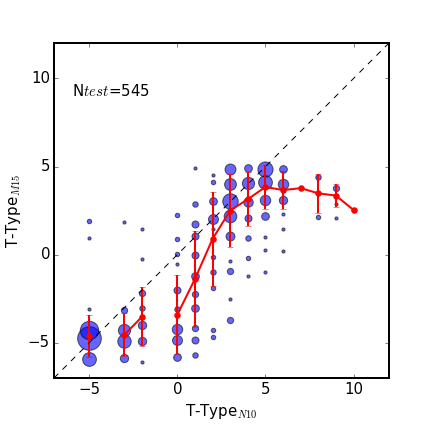}\par 
     \includegraphics[width=\linewidth]{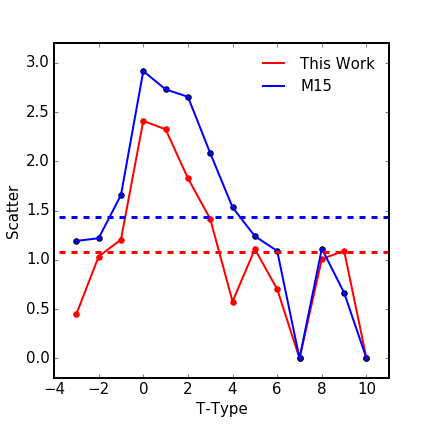}\par 
    \end{multicols}
\caption{ Comparison of our T-Type classification with the N10  (left panel) and  M15  (central panel). To better visualise it, we plot average binned values, where the size is proportional to the number of objects in each bin. The red dots show the mean value at each T-Type, while the error bars show the scatter. The right panel shows the scatter as a function of T-Type for our and M15 classifications. Our classification scatter is always smaller than the M15 classification one at all T-Types (except for T-Type=10) and on average $\sigma$=1.1  (red dashed line), comparable or even smaller than visual classification uncertainties ($\gtrsim$  1.3, \citealt{Naim1995}). } \label{fig:TType}
\end{figure*}

As stated in Section \ref{sect:data}, the N10 is a very detailed visual morphological catalogue which assigns an integer number to each galaxy (from -5 to 10) following a structural sequence. The detailed class for each number can be found in Table 1 of N10, but in short, T-Type < 0 correspond to ETGs, T-Type > 0 are spiral galaxies (from Sa to Sm), T-Type=0 are S0, while T-Type=10 are irregular galaxies.

We use 10000  galaxies with flag=0 (i.e., certain classification) for  training our T-Type models. We apply, though, a minor modification: in N10 the T-Type  minimum value is -5, but there are no galaxies defined as -4 or -1. To facilitate the model to fit a linear regression, we fill those gaps, so our  T-Types range from -3 to 10. The 0 still corresponds to S0/a, meaning that negative  T-Types correspond to early-type galaxies (E, S0-), positive T-Types correspond to spiral galaxies (from Sa to Sm) and 10 to irregulars. In this case, we use  \texttt{mean squared error (mse)}  as the loss function, which is widely used for linear regression algorithms.

\begin{figure*}
\centering
\includegraphics[width=0.9\textwidth]{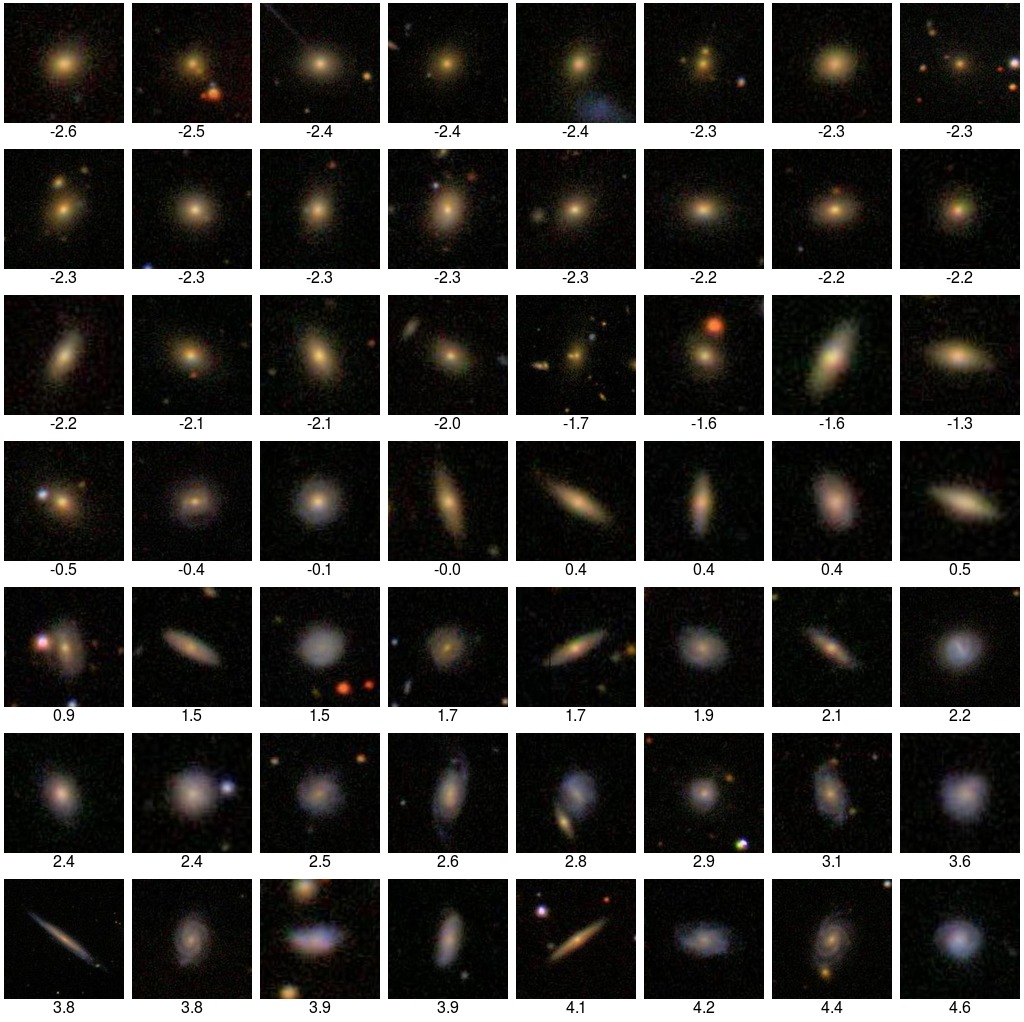}
\caption{\label{fig:T-Type-examples} Examples of galaxies sorted by the T-Type given by our model (shown in each cutout).}
\end{figure*}

In Figure \ref{fig:TType} we show the comparison between the classification obtained with our models and the N10 classification for a test sample of  $\sim$ 500 galaxies not used for training the model.
 The two classifications show an excellent agreement, with a median offset of b=0.03  up to T-Type $\leq$ 6. At higher values, the statistic is very scarce ($\leq$ 1\% ) and the model fails to converge. As a comparison, we show, for the same test galaxies, the T-Type obtained following equation 7 from M15, which transforms the probability values of being E, Sa, Sb or Sc, derived by \citet{Huertas2011} using support vector machine models, into a continuous T-Type sequence.  In this case, there is a median offset larger than one T-Type (b=1.7).  In this plot, we use the original T-Type value from N10 because equation 7 in M15 was optimised for the original catalogue. The scatter for our classification is on average $\sigma$=1.1,  comparable to or even smaller than expert classifier inter- comparisons \citep{Naim1995}. The scatter values are always smaller than the scatter for M15 classification for T-Type $\leq$ 6 ($\sigma$=1.4 on average, right panel Figure \ref{fig:TType}). Therefore, we consider that this is an improved T-Type catalogue compared to similar available catalogues, both in terms of accuracy  and number of classified galaxies ($\sim$ 50 times larger than the N10). In Figure \ref{fig:T-Type-examples} we show random examples of galaxies sorted by the T-Type derived with our models. The galaxies follow a smooth transition from E to spiral morphologies, as expected.

\subsection{Ell versus S0 models}
\label{sect:Ell/S0}

  \begin{figure*}
 \centering
 \includegraphics[width=0.9\textwidth]{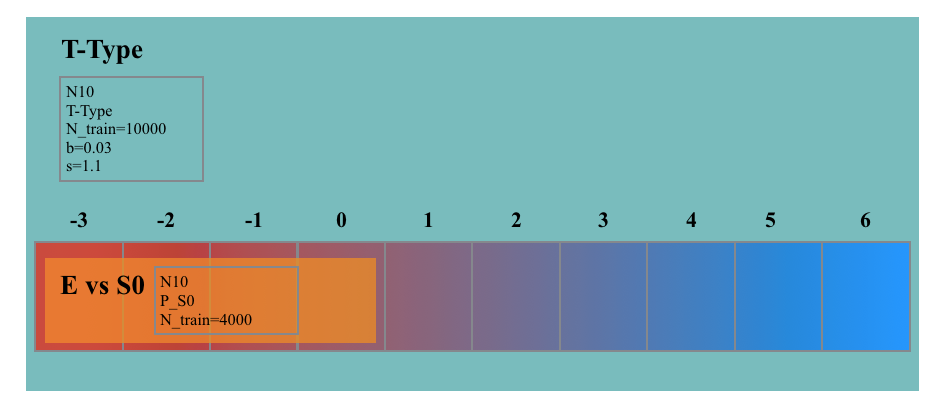}
 \caption{Scheme for our classification of T-Type questions. The main turquoise box represents the model for obtaining a T-Type value with some characteristics framed in grey (from top to bottom: the catalogue used for training,  the output of the model, the number of galaxies used in the training, the average bias and scatter). The coloured smaller boxes (red to blue) show the reliable T-Type outputs (at higher T-Types the statistic is scarce and our models deviate from the expected values).  An additional model to enable a distinction between pure E and S0 galaxies is represented as an orange box, and is only meaningful for galaxies with T-Type $\leq$ 0.}
 \label{fig:scheme2} 
 \end{figure*}

The performance of our model is excellent for the intermediate T-Types. However, it shows some flattening at  the edges (see Figure \ref{fig:TType}). For T-Type > 6 that is obviously due to insufficient statistics. On the other hand, the model trained to distinguish between such different morphological types as  spirals or ellipticals, is not able to clearly separate between  pure E and S0 galaxies, which share many characteristics.  In fact, 70\% of galaxies classified as ETG (T-Type $\leq$ 0) are assigned a T-Type < -2 and the largest scatter  is  precisely found for T-Type=0. Given that we have enough ETGs to provide a more accurate classification, we train an additional model to separate E  from S0 galaxies. We select galaxies with input T-Type $\leq$ 0 (and flag=0) and label as positive examples those with -3  $\leq$  T-Type $\leq$ 0 (S0-, S0, S0+ and S0/a, as defined in Table 1 form N10) and as negative those with T-Type=-5 (c0, E0, E+). We train the model with 4000 galaxies loading the weights of  the T-Type model, i.e., the  weights are initialized to the value learned by the CNN trained for the T-Type classification described  in Section \ref{sect:T-type model}. The  model output is $P_{S0}$, i.e., the probability of being S0 rather than E. A schematic classification for the models presented in this Section is shown in Figure \ref{fig:scheme2}.

\begin{figure}
\centering
\includegraphics[width=0.51\textwidth, trim=2cm 0 0 0]{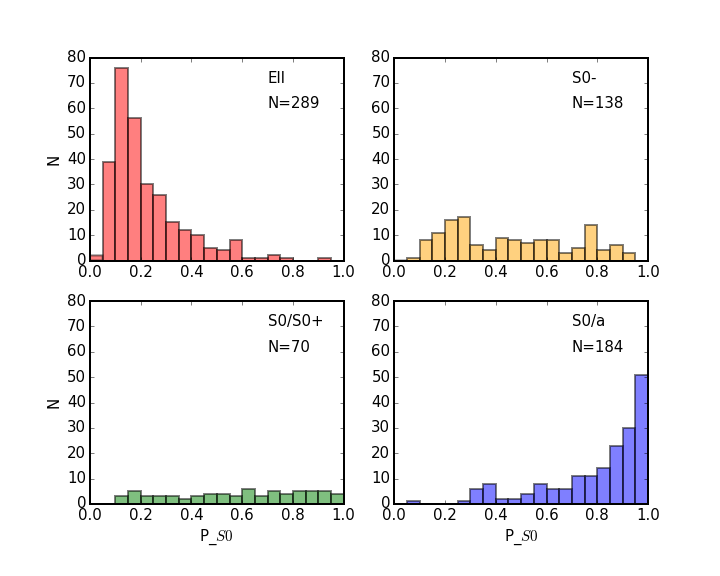}
\caption{ Distribution of the probability of being S0 rather than E (P$_{S0}$)  obtained with our model for a test sample of ETGs divided in 4 classes (according to N10): E, S0-, S0/S0+ and S0/a. The number of galaxies of each class is shown in each panel. The distribution is clearly skewed towards low $P_{S0}$ for the pure E galaxies and towards higher values for the S0/a. For the intermediate classes, the distribution spans over almost the whole probability range, as expected.} 
\label{fig:Ell-S0}
\end{figure}

To test this model, we apply it to a sample of 681 galaxies  not used in the training with T-Type $\leq$  0 and study the  $P_{S0}$ distribution for each ETGs sub-sample (Figure \ref{fig:Ell-S0}). The model is very efficient at identifying pure ellipticals: only 6\% of the test sample with T-Type = -5 in N10 is assigned  $P_{S0}$ > 0.5. Most of the S0/a are also correctly assigned a high $P_{S0}$, although there is a 10\% of them for which  $P_{S0}$ < 0.5. For the intermediate types (S0-, S0 and S0+), the $P_{S0}$ spans over all $P_{S0}$ values, as expected. We do a complementary check by comparing our $P_{S0}$ values with the bulge classes  (BC) values of the \citet{Cheng2011} catalogue (described in Section \ref{sect:data-test}) for a sample of $\sim$600 galaxies in common. We find that 95\% of galaxies with BC=3 (corresponding to Sa) have  $P_{S0}$ > 0.5, while only 11\% of galaxies with BC=1 (corresponding to E)  have  $P_{S0}$ > 0.5. The fraction of BC=2 galaxies with  $P_{S0}$ > 0.5 is 62\%, as expected for an intermediate class. Our classification presents  larger purity and completeness values when compared to the visual classifications from \cite{Cheng2011} than their automated classification method (75\% completeness and  73\% purity for the bulge identification, 83\% completeness and 70\% purity for the disks). When compared to the automated classification provided in \citet{Cheng2011}, our classification is not so accurate: 25, 56 and 84\% of galaxies with BC=1, 2 and 3, respectively have $P_{S0}$ > 0.5. This is an indication of our model being more efficient in distinguishing between E and Sa than the automated classification presented in \citet{Cheng2011}.

We conclude that this model efficiently allows distinguishing between pure E and Sa galaxies, which is a subtle classification task even for astronomers. We caution the reader that, although we provide a $P_{S0}$ value for each galaxy in our catalogue, it should only be used for galaxies with T-Type $\leq$ 0, for which the model was trained.

\subsection{Barred galaxies}
\label{sect:bar-nair}

\begin{figure}
\centering
\includegraphics[width=0.49\textwidth, trim=0 0 0 0]{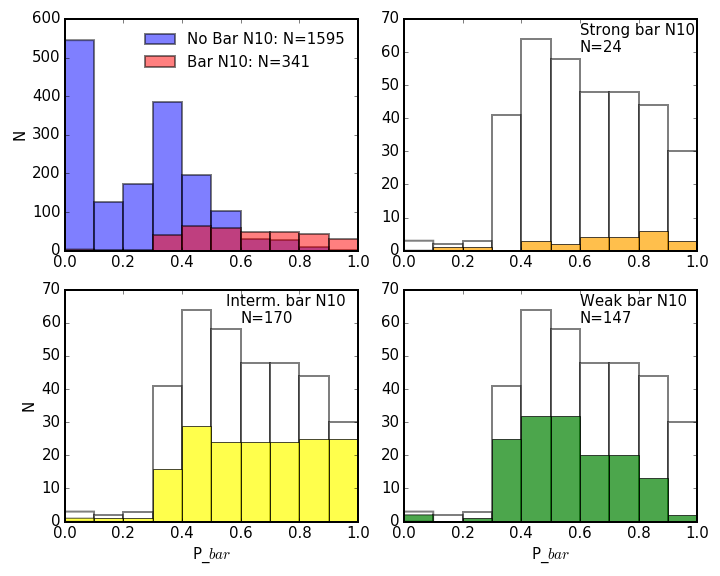}
\caption{Distribution of the probability of having bar signature, P$_{bar}$,  given by our N10 based model   for a test sample divided in 4 classes (according to N10): no bar, strong bar, intermediate bar and weak bar (filled histograms in blue, orange, yellow and green, respectively). In the upper left panel, we show  the  P$_{bar}$ distribution  for the barred (red) and unbarred (blue) galaxies. In the other panels, we show the P$_{bar}$ distribution for the barred galaxies (white empty histogram) and for the different classes of barred test galaxies (filled coloured histograms). The number of galaxies of each class is shown in each panel. The values are clearly skewed towards low values for the unbarred sample and towards higher values for the strong and intermediate bars. For the weak bars, the distribution peaks around  P$_{bar}$ $\sim$ 0.5.}
\label{fig:bar-nair}
\end{figure}

\begin{figure}
\centering
\includegraphics[width=0.5\textwidth]{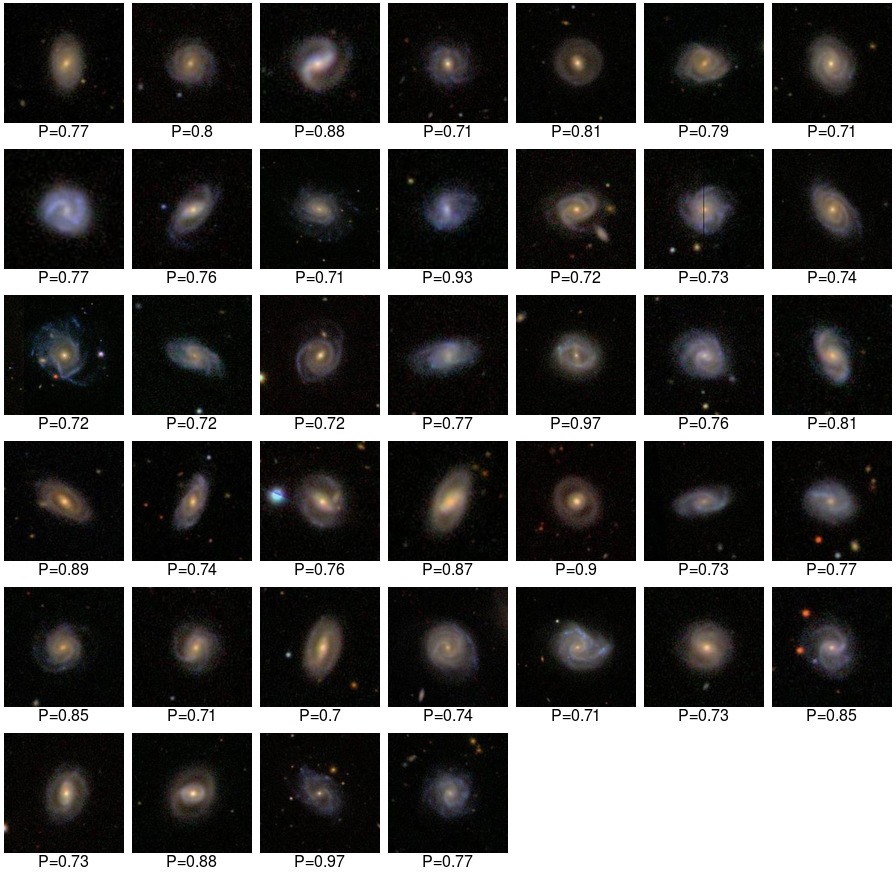}
\caption{ The 39 FP cases  in our N10-based bar classification. These are galaxies classified as unbarred by N10 but for which our model predicts $P_{bar}$ > 0.7. In most of the cases, there is a significant central structure, which could be considered as a bar or strong central bulge plus spiral arms. No catastrophic failures are found. }
\label{fig:FP-bar-nair}
\end{figure}

The N10 catalogue includes, in addition to the T-Type, a detailed visual classification of bars, divided into different classes - strong, intermediate, weak, etc. We take advantage of their bar classification to train an alternative model to the GZ2 based for barred galaxies. We focus on this particular characteristic because our GZ2 bar model is the one with the worst results (see Section \ref{sect:GZ2-models}).  In addition, the GZ2  bar classification is only efficient identifying  the strongest bars, while half of the galaxies with weak bar signatures are missed, as already mentioned in  Section \ref{sect:GZ2}.

We select a sample of barred galaxies (Bar flag > 0 from N10) and non-barred (Bar flag = 0). We train our model using 7000 galaxies,  of which approximately 20\% are barred. We load the weights from the GZ2 bar model, i.e., the  weights are initialised to the value learned by the CNN trained for the GZ2 bar classification described in  Section \ref{sect:train}. The results of applying the model to a test sample not used in the training, including 1595 unbarred galaxies and 341 barred galaxies,  is shown in Figure \ref{fig:bar-nair}.   We plot the $P_{bar}$ distribution of our model for galaxies belonging to those four different classes.  We correctly classify 90\% of unbarred galaxies ($P_{bar}$ < 0.5, TN) and 80\% of strong bar galaxies ($P_{bar}$ > 0.5, TP). However, the scarce number of strong bars (24) makes the statistics very noisy and there are actually  5 FN, of which only 2 have  $P_{bar}$ < 0.4. We visually checked those galaxies, finding that the two extreme cases ($P_{bar}$ < 0.3) were affected by close neighbours.  Setting $P_{thr}$ > 0.4, we obtain 88 and 80\% of TP for the intermediate and weak bar samples, respectively. The $P_{bar}$ distribution for the weak sample takes smaller values than the stronger bar sample,  indicating that our $P_{bar}$ could also be used as a proxy of bar strength. We visually inspected the FP cases, i.e.,  39 galaxies classified as unbarred in N10 but for which our model predicts $P_{bar}$ > 0.7 (shown in Figure \ref{fig:FP-bar-nair}). In most of the cases, there is a significant central structure, which could be considered as a bar or strong central bulge plus spiral arms. We conclude that no catastrophic failures are found.

As a complementary exercise, we study how well the model trained with the GZ2 catalogue  performs with respect to the N10 classification. The GZ2 bar model recovers  96, 80,  and 45 \% of the strong, intermediate and weak N10 bars.  Note that the model trained with GZ2 bar classification is even more efficient in identifying the galaxies with the strong bar signatures, but it fails to recover the weak ones. This demonstrates how the Deep Learning models  are  affected by the training sample. 

\section{Comparing this catalogue with the Galaxy Zoo 2}
\label{sect:catalogue}

In this Section, we summarise the content of the catalogue released with this paper and compare it with the GZ2 catalogue.

\begin{table}
\centering
\begin{tabular}{l | l | l| c |}
Col.  & Name & Meaning  & Train sample \\
\\\hline
1 & dr7objid &  SDSS ID  & \\
 2 & galcount & Meert15 ID &  \\
3 & $P_{disk}$ & Prob. features/disk  & GZ2\\
4 & $P_{edge-on}$ &  Prob. edge on & GZ2\\
5 & $P_{bar-GZ2}$&  Prob. bar signature &  GZ2\\
6 & $P_{bar-N10}$ &  Prob. bar signature &  N10 \\
7 & $P_{merg}$ &  Prob.  merger & GZ2 \\
8 & $P_{bulge}$ & Prob. bulge prominence & GZ2\\
9 & $P_{cigar}$  & Prob. cigar shaped & GZ2\\
10 & T-Type  &  T-Type & N10  \\
11 & $P_{S0}$  & Prob. S0 vs E &  N10\\\hline
\end{tabular}
\caption{\label{tab:catalogue} Content of the catalogue released with this paper. The catalogue contains 670,722 rows, each corresponding to a galaxy from the M15 sample. The last column of this table indicates which catalogue has been used for training each model.}
\end{table}

Table \ref{tab:catalogue}  summarises the content of the catalogue presented in this work.  A detailed explanation on the training procedure and the performance of all the models  has been presented throughout the paper.  The catalogue includes classification values for all 670,722 galaxies from the \cite{Meert2015} sample, as explained in Section \ref{sect:sample}.  We provide a probability value for each question and galaxy. Depending on the user purpose, a $P_{thr}$ value should be chosen to select \textit{positive} examples.  Values of precision ($\sim$ purity) and TPR ($\sim$ completeness) for three  $P_{thr}$ values  are tabulated in Table \ref{tab:PRThr} with this objective. For example, if one  aims to select a very pure \textit{edge-on} sample, $P_{thr}$ $\sim$ 0.85 would be a good choice, while for \textit{disk/feature} galaxies $P_{thr}$ $\sim$ 0.50 would be enough.

This is the largest and more accurate morphological catalogue available for the SDSS data up to date. Once trained, applying the models to images of galaxies without any previous classification is straightforward and no time-consuming. Therefore, our catalogue contains a homogeneous GZ2-type classification for a sample of galaxies twice as large as the W13 catalogue. It also provides a T-Type value for a sample of galaxies 50 times larger than the previous available T-Type catalogue (N10) and a finer separation between E and S0 galaxies. This is the first time, to the best of our knowledge, that a T-Type and an E/S0 classification are obtained with Deep Learning algorithms.

\begin{figure}
\centering
\includegraphics[width=0.49\textwidth]{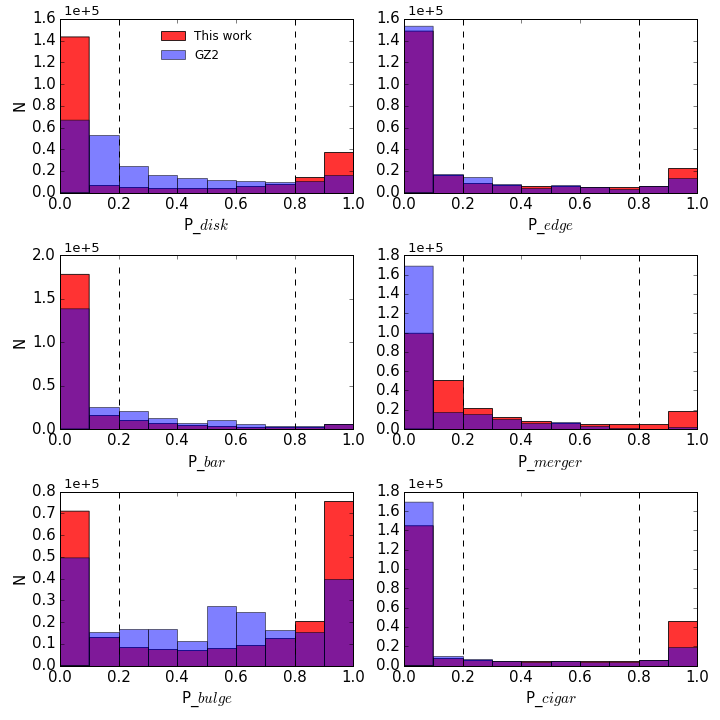}
\caption{ Probability distributions of our models (red)  compared to the original GZ2 probabilities (blue) for each of the GZ2-type classification tasks presented in our catalogue, for the  sample of galaxies in common (233,472). The dashed vertical lines mark the probability threshold which translates into $a(p)$ > 0.3 (for binary classifications). The   GZ2 P$_{bulge}$ value is  the sum P$_{dominant}$ + P$_{obvious}$, the same we use for  training our models (see section \ref{sect:2-answ}). Note that the P$_{merg}$ comparison is not straightforward due to the different approach used in our models with respect to the GZ2 decision tree (see \ref{sect:GZ2-models}). }
\label{fig:PDF}
\end{figure}

The probability distributions of our models are compared with the GZ2 ones  in Figure \ref{fig:PDF}.  We recall that we use a \texttt{sigmoid} activation function for our binary classification models. This function tends to bring the output values to either end of the probability distribution (0 or 1). In addition,   by training the models with robust examples, the machine learns how to recognise the features and the output probabilities. This causes our probability distributions to be generally more  bimodal for most of the tasks.  Our probabilities should be more objective in the sense that they measure similarity to robustly classified objects. Having a bimodal probability distribution is helpful because it removes galaxies with intermediate probabilities - low $a(p)$-, which are difficult to interpret for scientific purposes. This is very evident for Q1, where the fraction of galaxies with $a(p)$ > 0.3 increases from 56\% for the GZ2 to 86\% for our catalogue. The comparison of the fraction of galaxies with a  \textit{certain} classification for the questions belonging to the second or third tier of the GZ2 tree, such as $P_{bar}$, P$_{bulge}$ or  P$_{merger}$ is more complicated due to the thresholds for determining well-sampled galaxies in GZ2 (according to Table 3 in W13).  On the other hand, there are tasks, such as  P$_{edge}$ or P$_{cigar}$, which show similar distributions for both the GZ2 and our model.

\begin{figure}
\centering
\includegraphics[width=0.48\textwidth, trim={1cm 0 0 0}]{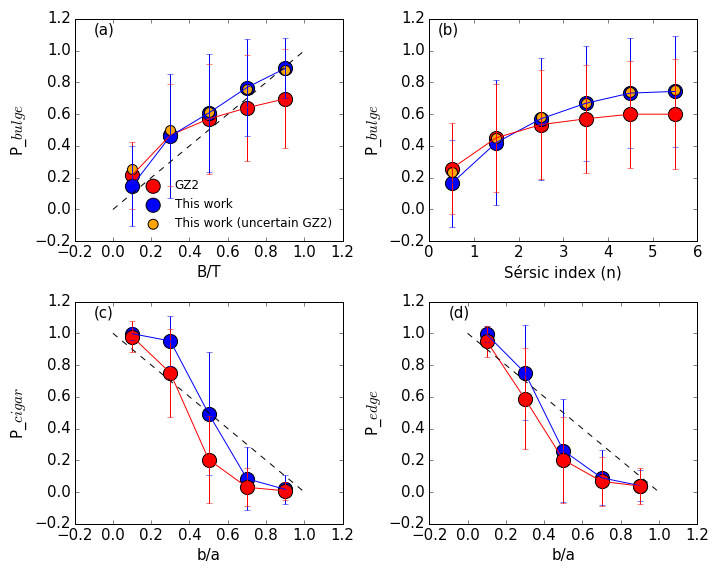}
\caption{\label{fig:Prob-params}  Mean probability values for our models (blue) and the GZ2 catalogue (red)  in bins of  morphological parameters  (extracted from the S\'ersic-exponential  photometric   catalogue presented in M15) for the sample of 233,472 galaxies in common with the GZ2 catalogue. The error bars represent the standard deviation in each bin. (a)  P$_{bulge}$ versus bulge-to-total ratio; (b)   P$_{bulge}$ versus S\'ersic index; (c) P$_{cigar}$ versus ellipticity; (c) P$_{edge}$ versus ellipticity. The orange dots  in panels (a) and (b) show the mean $P_{bulge}$ distributions according to our catalogue for galaxies with  0.2 > $P_{bulge-GZ2}$ > 0.8. The existence of a similar correlation for this subset of galaxies for which the GZ2 classification is \textit{uncertain} is an indication of the quality of our classification.}
\end{figure}

A test on the reliability of the output probabilities of our models is  their correlation with other morphological parameters. As a reference, we use the morphological parameters provided in the S\'ersic-Exponential photometric catalogue presented in M15. In Figure \ref{fig:Prob-params} we show  mean probability values in bins of   bulge-to-total ratio ($B/T$), bulge S\'ersic index ($n$) or ellipticity ($a/b$). There is a clear correlation between  P$_{bulge}$ and $B/T$, as well as $n$. This correlation is stronger for the probabilities  provided by our catalogue than for the GZ2 values,  demonstrating the physical meaning of the output probabilities of our models.  The correlation between $a/b$ and $P_{cigar}$ or $P_{edge}$ is also very evident, for both our probabilities and the  GZ2 ones. This is expected, given the similar probability distributions for these two tasks for the GZ2 and our models (see Figure \ref{fig:PDF}). We also show the mean  $P_{bulge}$ according to our model in bins of $B/T$ and $n$ for the sub-sample of galaxies with low $a(p)$ in the GZ2 catalogue (i.e.,   0.2 > $P_{bulge-GZ2}$ > 0.8). The correlation is also evident for this sub-sample, demonstrating that our probabilities have a physical meaning even for the galaxies with uncertain GZ2 classifications.

\begin{figure}
\centering
\includegraphics[width=0.5\textwidth]{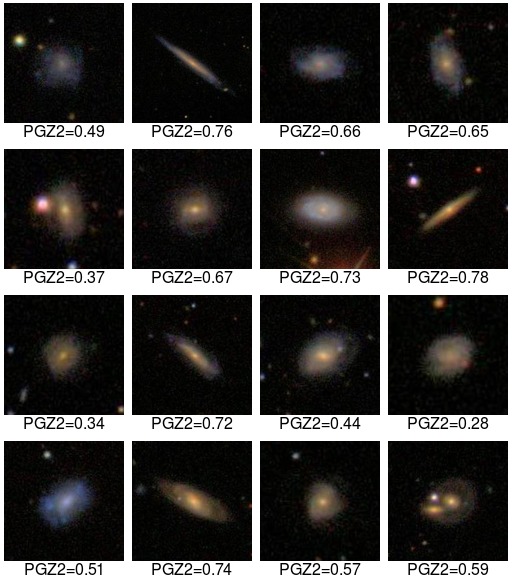}
\caption{\label{fig:disk-examples}  Examples of  galaxies with high probability of having disk/features by our model  (Q1, $P_{disk}$ > 0.9) but with uncertain  GZ2 classification ($a(p_{GZ2})$ < 0.25). The number shown in the cutouts is the probability given by the GZ2 catalogue. }
\end{figure}

\begin{figure}
\centering
\includegraphics[width=0.5\textwidth]{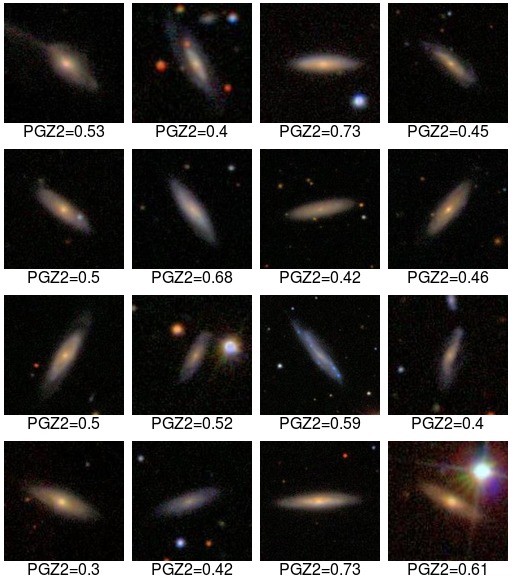}
\caption{\label{fig:edge-examples}  Examples of galaxies with high probability of being edge on (Q2, $P_{edge-on}$ > 0.9) by our model and  an uncertain  GZ2 classification ($a(p_{GZ2})$ < 0.25). The number shown in the cutouts is the probability given in GZ2 catalogue.}
\end{figure}

\begin{figure}
\centering
\includegraphics[width=0.5\textwidth]{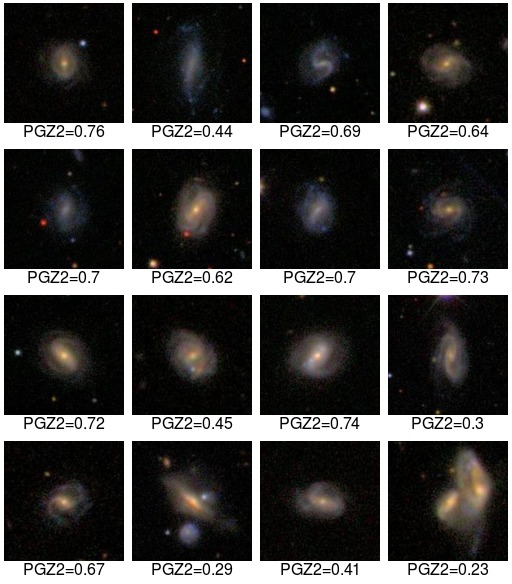}
\caption{\label{fig:bar-examples}  Examples of galaxies with high probability of having bar signatures (Q3, $P_{bar}$ > 0.9) by our model and an  uncertain  GZ2 classification ($a(p_{GZ2})$ < 0.25). The number shown in the cutouts is the probability given in GZ2 catalogue.}
\end{figure}

Unfortunately,  there is no  quantitative way to demonstrate that our classification works better than GZ2 for galaxies with low $a(p_{GZ2})$, since there is no ``true reference" catalogue. We can only test our models by visual inspection. In Figures \ref{fig:disk-examples}, \ref{fig:edge-examples} and \ref{fig:bar-examples} we show arbitrary examples of galaxies with high output probabilities  from our models (P > 0.9) and low $a(p_{GZ2})$. In most of the cases, the classification given by our model is robust and correct, while the GZ2 probabilities are much lower (and thus, uncertain).


We also checked the number of \textit{catastrophic errors}, defined as galaxies for which GZ2 and our classification are very different.  The fraction of galaxies for which our model predicts P > 0.8 and GZ2 has P < 0.4 (or vice versa) is 2.5, 1.7 and 1.9\% for Q1, Q2, Q3, i.e., less than 3\% for all questions with two possible answers in GZ2 decision tree. For Q6, mergers, the discrepancy is a bit larger (7.2\%), but we want to stress the difficult comparison between our model and GZ2 for this particular question. 

An advantage of our catalogue with respect to the GZ2 is that our probabilities are not affected by the number of votes (i.e., the number of volunteers who  who answered a particular question).  Note,  however, that the minimum five vote requirement for the training sample  of each GZ2 task means that the models somehow contain selection effects of previous questions of the classification scheme (Figure \ref{fig:scheme}).  For example, the  probabilities of being bar or edge-on have been  trained with galaxies which at least five volunteers have classified as \textit{disk/features}. Therefore, the probability of a question contained in an upper-level box of Figure \ref{fig:scheme} should only be completely trusted for positive examples of that particular answer. 

Care should also be taken with the $P_{S0}$ value, whose meaning is to differentiate  E from S0, i.e., should only be applied to galaxies with T-Type $\leq$ 0. We also advise that, due to the limited merger examples,  $P_{merger}$ has difficulty in selecting real on-going mergers. After visual inspection, the P$_{merg}$ value looks like a better proxy to clustered galaxies or projected pairs than to  actually interacting galaxies. For simplicity in the catalogue construction and release, we provide one value for each  question and galaxy, but we caution  the user to properly understand the meaning of each probability when using it.

\section{Conclusions}
\label{sect:conclusions}

In this work, we present a morphological catalogue for a sample of $\sim$ 670,000 galaxies from the SDSS DR7  corresponding to the sample analysed by \cite{Meert2015,Meert2016}. The morphological classifications are obtained with Deep Learning algorithms using CNNs, and the models are trained with the best available visual classification catalogues \citep{Nair2010, Willett2013}.

We use the GZ2 catalogue presented in W13 to train GZ2 classification models: presence of a disk or features, edge-on disks, bar signature, roundness, bulge predominance and merger signature. The main novelties of our training approach with respect to previous works (e.g. \citealt{Dieleman2015}) are:
\begin{itemize}
\item we independently train each question from the GZ2 scheme listed in Table \ref{tab:questions};
\item we use in the training only galaxies with small GZ2  classification uncertainties (large agreement,  $a(p)$ $\geq$ 0.3,  between classifiers). This allows the models to easily extract the fundamental features for each question;
\item we train the questions in binary classification mode, i.e., only two answers (yes or no) are allowed for each question. The output of each model is the probability of being a positive example, as shown in Figure \ref{fig:scheme}, and takes values between 0 and 1.
\end{itemize}

Our models show large accuracy values (> 97\%) when tested against  a sample with the same characteristics as the one used in the training (i.e., \textit{robust} GZ2 classifications). There is a P$_{thr}$ value for each question  for which both TPR ($\sim$ completeness) and precision ($\sim$ purity) are > 90\% (except for the bar sign, for which TPR and precision only reach $\sim$ 80\%, see discussion in section \ref{sect:GZ2-models}). These values are listed in Table \ref{tab:PRThr}. Our morphological catalogue includes  a homogeneous classification for 670,722 galaxies, increasing by a factor $\sim$ three the statistics with respect to GZ2. In addition, we obtain a more unambiguous classification for some of the GZ2-type tasks (see Figure \ref{fig:PDF}). This result is particularly important regarding the question  about the presence of \textit{disk/features}, where the number of galaxies with   \mbox{$a(p)$ > 0.3} increases from 56\% in the GZ2 to 86\% in our catalogue. 

We complement the GZ2 type classification with a \mbox{T-Type} value. To this purpose, we train the models with the visual classification catalogue presented in N10. The catalogue presented in this paper  is the first T-Type classification obtained with CNNs - to the best of our knowledge - and represents a significant increase in terms of statistics compared to previous available T-Type catalogues ($\sim$50 times larger than the N10 catalogue).  In this case, we train the model  using a regression mode, so the output ranges from -3 (E) to 10 (irregular). As shown in Figure \ref{fig:TType}, when compared to the T-Type from N10, our classification shows no offset and a scatter comparable to or even smaller than typical expert visual classifications (b=0.03, $\sigma$=1.1).  These values are smaller than the ones obtained when comparing N10 T-Type with the classification proposed by M15 (b=1.7 and $\sigma$=1.4). We provide an additional model enabling a separation between E and S0 galaxies. This classification is tested against the N10 and \cite{Cheng2011} catalogues with a great success rate (94\% of TP pure  E galaxies when compared to N10, see Figure \ref{fig:Ell-S0}). We also use the N10 bar classification to obtain an alternative model to the GZ2 based for the bar signature, in order to have a complementary bar indicator to the GZ2 based, with a high success rate (> 80\% TP and TN, see Figure \ref{fig:bar-nair}).

We remind that applying the models to images of SDSS galaxies without any previous classification is straightforward and no time-consuming. Therefore, in a forthcoming work (Dom\'inguez S\'anchez et al. in preparation), we plan to complement the morphological classification catalogue by applying the models  to other SDSS samples, such as the MaNGA dataset \citep{Bundy2015}.

\section*{Acknowledgements}

The authors would like to thank the anonymous referee  for comments that improved the paper. This work was funded by the French National Research Agency (ANR) project ASTROBRAIN (P.I. MHC) and by UPenn research funds (MB). The authors are also thankful to Google for the gift given to UCSC (``Deep-Learning for Galaxies") which has also greatly contributed to making this work possible. Finally, HDS and MHC would like to thank our UCSC colleagues Sandy Faber, David Koo, Joel Primack for very productive discussions which helped to improve the content of the paper. 




\bibliographystyle{mnras}
\bibliography{sample} 




\bsp	
\label{lastpage}
\end{document}